\documentclass{ws-p8-50x6-00}

\begin{document}

\title{WIMP searches with superheated droplet detectors: Status and 
Prospects}

\author{J.I. Collar$^{\S}$, D. Limagne, T. Morlat, J. Puibasset, G. Waysand}

\address{Groupe de Physique des Solides (UMR CNRS 75-88), 
Universit\'es Paris 7 \& 6,\\ 75251 Paris Cedex 05, France}

\author{T.A. Girard}

\address{$^{\S}$Centro de F\'\i sica Nuclear, 
Universidade de Lisboa, 1649-003 
Lisbon,Portugal}  

\author{H.S. Miley}

\address{Pacific Northwest National
Laboratory, Richland, WA 99352,
USA}  


\maketitle

\abstracts{
SIMPLE ({\underline S}uperheated {\underline I}nstrument 
for {\underline
M}assive {\underline P}artic{\underline {LE}} searches) 
employs superheated droplet detectors (SDDs)
to search for Weakly Interacting Massive Particle (WIMP) dark matter.
As a result of the 
intrinsic SDD insensitivity to minimum 
ionizing particles and
high fluorine content of target liquids, competitive WIMP 
limits were already obtained at the early prototype stage. We 
comment here on the 
expected immediate increase in sensitivity of the program and on 
future plans to exploit this promising technnique.}

\section{Introduction}
Superheated Droplet Detectors (a.k.a. SDDs or Bubble Detectors) 
consist of a dispersion of small drops (radius $\sim10~\mu m$)
of superheated liquid in a gel or viscoelastic matrix, with a droplet 
filling factor of few \%. This relatively recent 
development in radiation detection technology \cite{apfel1} 
combines the 
well-known characteristics of bubble chambers with several 
advantageous features: the SDD matrix
isolates the fragile metastable system from vibrations and 
specially
from convection
currents (damped in gels), while the smooth
liquid-liquid interfaces
impede the continuous triggering on surface
impurities
that occurs 
even in ``clean''
bubble chambers. In this way, the extended lifetime
of the superheated state
allows for new practical applications such as 
personnel and area neutron dosimetry.
In the moderately superheated
industrial refrigerants used in SDDs,
bubbles can be produced only by
particles having elevated
stopping powers, as is the case for low-energy
nuclear recoils. This behavior is 
described by the
``thermal spike'' model \cite{seitz}: for 
the transition to occur, a vapor
nucleus or ``protobubble'' of radius $> 
r_{c}$ ($\sim$tens of nm) must be created, while only the 
energy deposited along a distance
comparable to this critical radius $r_{c}$
is available for its formation.
Hence, a double
threshold is imposed: the requirement that the 
deposited energy
$E$
be larger than the thermodynamical
work of formation of the
critical nucleus, $E_{c}$, and that this
energy be lost by
the particle
over a distance $\textrm{O}\!\left(r_{c}\right)$,
i.e., a minimum stopping
power. Protobubbles formed by energy depositions not meeting 
both demands 
shrink back to zero; otherwise,
the whole droplet vaporizes producing a characteristic 
sound\footnote{This sharp, high-pitched sound is similar to that produced by 
spraying water on hot oil, the physical mechanism (vaporization 
of superheated water) being the same.} used for detection purposes.  
Both thresholds depend on
operating conditions\cite{prletc22,njp}: keV nuclear recoils are 
detectable at room temperature and atmospheric pressure, 
allowing for a low-cost WIMP search free of the complications 
associated to cryogenics.
Most importantly, the minimum $dE/dx$ requirement 
provides a total insensitivity to most backgrounds interfering 
with WIMP searches, internal $\alpha$-emitters being the only concern\cite{myprd}. 
\section{Detector Fabrication}
SDDs of active mass O(1) kg can considerably
extend the present experimental
sensitivity  to WIMPs well into the
region where new supersymmetric
particles are expected \cite{myprd}. The modest active
mass of commercially available SDDs
($\sim0.03$ g refrigerant/dosimeter), together with a desire 
to control
and understand the fabrication process,
lead us to develop a large-volume 
pressure reactor dedicated to
SDD production. Three years of R\&D were necessary to produce  
SDD modules simultaneously meeting all requirements of alpha-emitter 
radiopurity, 
stability and large-mass (presently $\sim 1000$ times that of commercial SDDs) 
necessary for this new application. 
Details on the
fabrication process,  
purification and testing 
can be found in \cite{prletc22,njp}. The emphasis of the program has been 
on keeping the number of
detector components down to a minimum for reasons of radiopurity, 
safety and cost, imposing the need to solve multiple condensed-matter issues 
with as few elements as possible. 
The target mass of present
SIMPLE modules is limited only by the dimensions of the 
pressure reactor. They 
can be operated continuously for up to $\sim40$ d, 
while polyacrylamide gel-based commercial bubble dosimeters are intended for 
just a few hours of continuous exposure. 
The construction of recompression 
chambers that will allow for 
an indefinite exposure is underway. The material cost of the SDDs 
is very low, permitting a much
larger future
design.
\section{Calibrations}
The response of smaller SDDs to various
neutron fields has been
extensively studied \cite{harper,apfel3,derico}
and found to closely match theoretical
expectations. However, large-size, opaque SDDs
require independent
calibration:
acoustic detection of the explosion of the smallest 
droplets or of those most distant
from the piezolelectric acoustic sensors 
is not {\it a priori} guaranteed.  
Two separate
types of calibration have
been performed to
determine the target mass  effectively 
monitored in SIMPLE modules and to
check the calculation of the temperature- 
and pressure-dependent threshold energy 
$E_{thr}$
above which WIMP recoils can induce nucleations 
\cite{myprd,nagdy}. SIMPLE SDD modules have been exposed to a 
well-characterized $^{252}$Cf neutron
source at the TIS/RP calibration
facility (CERN) \cite{prletc22,njp}. 
A satisfactory agreement between the observed 
nucleation rate and Monte Carlo simulations was found at different 
pressures and temperatures \cite{prletc22,njp}. As a result of this calibration, the 
sound detection efficiency with present sensors was found to be 
$34\pm 2\%$ ($74\pm 4\%$)  at $P\!=$2 atm ($P\!=$1 atm), a decisive 
piece of information to obtain dark matter limits. A second type of 
calibration can be performed by homogeneously diluting 
a liquid $^{241}$Am source
(an alpha-emitter)
into the matrix prior to gel setting. The 5.5 MeV alphas and 91 keV 
recoiling $^{237}$Np daughters cannot induce nucleations at 
temperatures below $T_{\alpha}$ and $T_{\alpha r}$, respectively 
\cite{myprd,prletc22,njp}. These $^{241}$Am calibrations 
allowed us to corroborate the calculation of $E_{c}$ 
for each refrigerant (a good knowledge of $E_{c}$ is crucial to
determine the expected
SDD sensitivity to WIMP recoils in
different
operating conditions). An additional test of SDD response to 
low-energy nuclear recoils is in preparation
at the Sacavem (Lisbon) experimental reactor, using 
highly monochromatic (filtered) keV neutrons.
\section{First Results}
\begin{figure}[t]
\begin{center}
\epsfxsize=17pc 
\epsfbox{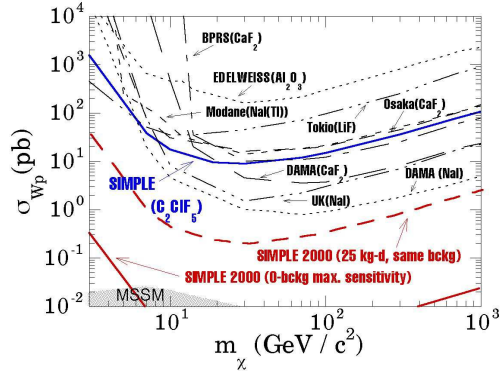} 
\end{center}
\caption{95\% C.L. limits on $\sigma_{Wp}$ 
extracted
from 0.19
kg-day of SDD exposure, compared with other experiments (DAMA 
limits stem from a $\sim 1.5\cdot 10^{4}$ kg-day exposure). PICASSO$^{16}$ 
has reported similar first limits. 
The lowest lines indicate the expected sensitivity of SIMPLE-2000 after 
25 kg-day, due to the increase in exposure only 
(dashed line) or at the maximum reachable level at this stage
(zero
background, solid
line off the scale).
``MSSM'' marks the tip of the region where a
lightest supersymmetric
partner is expected. \label{fig:radish}}
\end{figure}
The installation 500 m underground of SIMPLE modules started in July 
1999.
A decommissioned nuclear missile launching
control center has been converted into an
underground laboratory
\cite{wwwrustrel,geo}, facilitating this and other
initiatives.
The characteristics of this site (microphonic silence, unique 
electromagnetic shielding of the halls) make it specially
adequate for rare-event searches. Modules are placed inside a 
temperature-regulated water bath, 
surrounded by three layers of sound and thermal 
insulation. The 700 l bath, which rests on a dual 
vibration absorber, doubles as neutron moderator. Events in the modules and in 
external acoustic monitors are time-tagged, allowing to filter-out 
the small percentage ($\sim\!15$\%) of signals correlated
to human activity in the immediate vicinity of the experiment.
The signal waveforms are digitally stored, but no event rejection based on 
pulse-shape considerations is performed at this stage (the 
sharply-resonant piezoelectric sensors presently employed will eventually 
be substituted by others with a flatter spectral response, allowing an 
unequivocal identification of the nucleation sounds). Accounting for sound
detection efficiency and a
62\% fluorine mass fraction in R-115 ($C_{2}ClF_{5}$),
limits on the
spin-dependent WIMP-proton cross section 
$\sigma_{Wp}$ (\frenchspacing{Fig. 1}) were 
extracted
from the first 0.19 kg-day of exposure of a  
prototype module \cite{prletc22,njp}.
Those limits, while already competitive, were
impaired
by the large statistical uncertainty associated to the 
short exposure. A considerable improvement (\frenchspacing{Fig. 1})
is expected with the already implemented expansion of the bath to 
accommodate multiple modules. SIMPLE 2000 aims 
at an exposure of $\sim 25$ kg-day by
replacing  the detectors (in batches of seven) every four to six 
weeks,
repeating this cycle several times.
A moderated Am/Be neutron source will be
used at the end
of each run to assess {\it in situ} the sound detection efficiency of
each module. Runs using refrigerant-free modules
showed that
{\it a majority} of the prototype events arose from pressure microleaks 
in the old caps,
able to stimulate
the piezoelectric sensor \cite{prletc22,njp}. Plastic module caps have
now been replaced by
a sturdier design\footnote{At the time of this 
writing, the level of spurious unrejectable acoustic noise in the 
modules has been reduced by a factor $>20$ with respect to the prototype 
data in \cite{prletc22,njp}. Pressure microleaks have been eradicated: 
the remaining source of noise ($<0.1$ 
events/module/day) originates in 
thermal convection currents in the larger bath able to mechanically disturb
microphone cables, changing their 
capacitance. A first stage of signal amplification next to the 
piezoelectric transducers and/or improved cabling will avoid this.}.
The expected maximum sensitivity
of SIMPLE 2000
can already start to probe the spin-dependent neutralino parameter space 
(\frenchspacing{Fig.
1}). As a next step, ten recompressible (i.e., permanent) modules comprising a 
total active mass of $\sim 250$ g are under construction. 
\begin{figure}[b]
\begin{center}
\includegraphics[width=.7\textwidth]{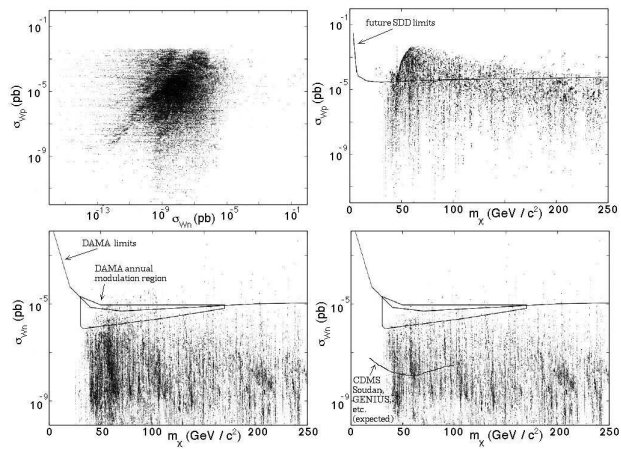}
\end{center}
\caption[]{A fluorine-rich SDD of active mass O(1) kg will be 
sensitive to neutralino WIMPs beyond the 
reach of the most ambitious planned cryogenic experiments (see text). 
}
\label{eps3.1}
\end{figure}
\section{Prospects}
Theoretically predicted 
spin-dependent (SD) WIMP scattering rates are generally smaller than their 
spin-independent (SI) counterparts, not profiting from a 
coherent enhancement in the cross section. Sensitive SD searches 
should 
therefore favor the use of fluorine-rich detectors, fluorine being 
by far the optimal SD target \cite{john}.
To stress the significance of the SD channel, we 
have recently emphasized \cite{njp} a not-so-obvious
complementarity
of SD and SI
searches in exploring the neutralino
phase space: the top-left frame in \frenchspacing{Fig.
2}
displays
points generated
with the help of NEUTDRIVER \cite{wimps}, 
each representing a
possible combination of MSSM parameters. The parameter space sampled
is the same as in \cite{bottino}. Only a weak
correlation between
the values of $\sigma_{Wp}$ (i.e., the SD coupling strength, corrected for local 
halo density) 
and $\sigma_{Wn}$ (idem for SI) is observed in the plot (note the 
scale).
As a result of this, a compact ($\sim 1$ kg active mass), fluorine-rich, 
low-background SDD
starting to
probe the neutralino $\sigma_{Wp}$ (top-right frame) generates a rather 
homogeneous rejection or ``cleaning'' of permissible MSSM models 
in a $\sigma_{Wn}$
exclusion plot (bottom-left frame,
without the constrains imposed
by the limits in top-right; bottom-right with them), 
down to values of $\sigma_{Wn}$
far beyond 
the reach of the most ambitious planned cryogenic WIMP searches.
Needless to say, the converse
can be stated of the way that 
sensitive SI searches will 
deplete the
$\sigma_{Wp}$
space and hence the complementarity of both approaches. A second 
observation that can be made from \frenchspacing{Fig.
2} is that the SD interaction determines the lower bound for 
the direct detection rate \cite{klapdor} (i.e., 
in most MSSM models $\sigma_{Wp}$ 
remains large enough for direct detection to be realistic, not 
necessarily the 
case for $\sigma_{Wn}$).

In view of this, if an exhaustive test of the neutralino-as-cold-dark-matter hypothesis 
is ever to be completed, the development of fluorine-rich 
detectors cannot be neglected: in this respect CFC-loaded SDDs represent an 
ideal opportunity. Nevertheless, the development of CBr$F_{3}$ and CI$F_{3}$ 
SDDs able to
simultaneously explore both SD and SI channels is another of our goals. 
PICASSO \cite{picasso} and SIMPLE intend to merge into an international 
effort (U. Montreal, U. Paris, U. Lisbon, Yale U., PNNL, SNO, 
Bose Institute) to fully exploit the promise of this approach 
to dark matter detection.

\end{document}